\documentstyle{article}
\newcommand{\myl}[1]{\label{#1}\eeq}
\newcommand{\myll}[1]{\label{#1}}
\newcommand{\myc}[1]{\cite{#1}}
\newcommand{\state}[2]{\begin{center}\vspace{\baselineskip}
\parbox{.85\textwidth}{{\bf #1:}\ {\em
#2}}\vspace{\baselineskip}\end{center}}
\newcommand{\myref}[1]{(\ref{#1})}
\newcommand{\mysref}[1]{Section\ \ref{#1}}

\newcommand{\beq}{\begin{equation}}
\newcommand{\eeq}{\end{equation}}
\newcommand{\beqa}{\beq\begin{array}{l}}
\newcommand{\fn}[1]{${}\sp#1$} 

\newcommand{\R}{${\bf R^4}$}

\newcommand{\Rx}{${\bf R^4_\Theta}$}
\newcommand{\mRx}{\bf R^4_\Theta}
\newcommand{\sx}{$\bf\Sigma^7_\Theta$}

\newcommand\xt{\times_\Theta}
\newcommand\dM{{\cal D}}
\newcommand\dMx{{\cal D}_\Theta}
\begin{document}
\begin{titlepage}
\begin{center}
{\Large \bf  Exotic Differentiable Structures\\ and \\
General Relativity}
\end{center}
\vspace{.5truein}
\begin{center}
{\large \bf Carl H. Brans\fn{1} and Duane Randall\fn{2}}
\end{center}
\vspace{.5truein}
\begin{center}
\fn{1}Physics Department\\ Loyola University\\ New Orleans, LA 70118  \\
e-mail:BRANS @ LOYNOVM.BITNET
\end{center}
\par
\begin{center}
\fn{2}Mathematics Department\\ Loyola University\\ New Orleans, LA
70118\par
\vspace{.5truein}
{\bf November 23, 1992}
\end{center}
\begin{abstract}
We review recent developments in  differential topology with special
concern for their possible significance to physical  theories, especially
general relativity.  In particular we are concerned here with the
discovery of the existence of non-standard (``fake'' or ``exotic'')
differentiable structures on topologically simple manifolds such as
$S^7$, \R and $S^3\times {\bf R^1}.$  Because of the technical
difficulties involved in the smooth case, we begin with an easily
understood toy example looking at the role which the choice of complex
structures plays in the formulation of two-dimensional vacuum
electrostatics. We then  briefly  review the mathematical formalisms
involved with differentiable structures on topological manifolds,
diffeomorphisms and their significance for physics.  We summarize the
important work of Milnor, Freedman, Donaldson, and others in developing
exotic differentiable structures on well known topological manifolds.
Finally, we discuss some of the geometric implications of these results
and propose some conjectures on possible physical implications of these
new manifolds which have never before been considered as physical models.
\end{abstract}
\end{titlepage}
\section{Introduction}\myll{Intro}

Recently there  have  been some  significant
breakthroughs in differential  topology  and  global  analysis  of
manifolds which may very well have considerable influence for space-time
models in physics.   We refer to  investigations of   the various
possible differentiable structures that can be put on
a given topological manifold.\footnote{There are equally intriguing
questions in the {\em homotopy} category, but for simplicity, we focus
here only on the topological one.}
A very natural, basic, but sometimes very difficult, question  is the
following:
\state{Question 1}{ If one smooth manifold is homeomorphic to another,
need it be diffeomorphic?}
Thus, if two manifolds have the same topology, must they necessarily have
the same (up to diffeomorphisms) differentiable structure?\footnote{We
emphasize that we are not referring to merely {\em different} structures,
but ones which are not {\em diffeomorphic}.  These points will be
discussed in more detail in \mysref{cstoy} and \mysref{dsm}.}  While this
question is clearly a fundamental one for mathematics, it must surely
also have physical content, since the basic model of space-time is that
of a smooth manifold, and diffeomorphisms are generally regarded
physically as re-coordinatizations, i.e., changing of (local) reference
frames.
As it  stands, Question 1 obviously is too general for much progress to
be made, so it is natural to restrict it to certain simple classes of
manifolds, specifically spheres, $S^n$, and Euclidean spaces, ${\bf
R^n}.$  \par
When restricted to spheres and restated as a conjecture,
\state{Conjecture 1}{Any manifold  homeomorphic to $S^n$ is  necessarily
diffeomorphic to this manifold with its standard differentiable
structure.}
  For dimension $n=1$ the  question is easy to resolve, with the answer
in the affirmative.  For higher dimensions, however, the problem becomes
more difficult.  A breakthrough was made in 1956 by Milnor\myc{miln} who
was able to provide a counter example by explicitly constructing
manifolds,  \sx, which are topological seven-dimensional spheres but
which are not diffeomorphic to the standard one.  This work  led to
further  classification results for such exotic differentiable structures
on higher dimensional spheres, reducing the problem to one of
homotopy\myc{KM}.  However, as of today, Conjecture 1 for four-spheres is
still an open question. \par
For Euclidean spaces, ${\bf R^n}$, the version would be
\state{Conjecture 2}{Any manifold  homeomorphic to ${\bf R^n}$ is
necessarily diffeomorphic to this manifold with its standard
differentiable structure.}
At first glance, it would seem that the topological triviality of ${\bf
R^n}$ would make this an easy matter to settle.  In fact, for
 $n=1,2,3,$ the question can be answered in the affirmative by
straightforward computational techniques, trivial for $n=1$, but more
arduous for $n=2,3.$  Indeed, for $n\le 3$, {\em any} topological n-
manifold can be given a differentiable structure, compatible with its
topology, unique up to diffeomorphism.  On the other hand, for $n\ge 5,$
Smale's  h-cobordism theorem provides a basic tool for settling the
question in the affirmative, for ${\bf R^n}$. However, these results
failed to give any insight for the $n=4$ special case.  To summarize, the
best that could be said until 1982 was
\state{Theorem 1}{Conjecture 2 is true for any $n\neq 4.$}
It is probably reasonable to say that most workers expected that a
similar affirmative result would ultimately be obtained for $n=4$.  After
all, the topology is trivial, so what reason would there be to suppose
that any non-trivial structure could be imposed on this space, when the
same cannot be done for any other dimension?
\par
Thus, it came as quite a surprise when the work of
Freedman\myc{fr} and Donaldson\myc{D} building on earlier results of
Casson, established   the
existence  of  an   ``exotic'' (``fake'' or ``non-standard'')
differentiable
structure on the topological  manifold  ${\bf R}^4$.  Shortly thereafter
Gompf was  able to show the existence of several \myc{G1}, and then even
an infinity\myc{G2}, of distinct (i.e. not diffeomorphic) structures on
${\bf R^4}.$ Freedman and Taylor\myc{FT} were even able to demonstrate
the existence of a universal exotic ${\bf R^4}$ in which all others could
be smoothly embedded.   Some reviews of the subject are
available\myc{surv}. \par
The existence of such exotic structures is a strikingly counter-intuitive
result.  It means that although each of these manifolds is topologically
equivalent to \R, there is no local coordinate patch structure in which
the global topological coordinates, ordered sets of four numbers, are
everywhere smooth.  Other strange effects occurring in these manifolds
will be discussed in \mysref{exds} below.
\par
The path to the discovery of such manifolds, which we denote generically
by \Rx, is much more circuitous and mathematically involved than for the
\sx.  The bad news then is that following the argument in detail requires
a great deal of mastery of many branches of mathematics.  The good news,
from our viewpoint, is that this wandering journey involves mathematical
excursions touching on such strongly physics-based topics as Dirac
spinors, moduli spaces of Yang-Mills instantons and even an intersection
form, $E_8$, identical to the Cartan form for the exceptional group
recently studied in superstring theory.
In view of the exceptional role played by the physically significant
dimension four, and of all of the input from physics-based tools to the
mathematical developments, it is somewhat surprising that the impact  of
these  mathematical
discoveries on physics has not been more widely  explored  in  the
general physics literature.  To date, mention of exotic structures
in  physics papers has been mainly confined to problems related
to quantization of gravity or supergravity.
For example,
Witten\myc{Witten} discusses them in the context of supergravity and
superstring theory in ten dimensions, although he argues that their
effects cancel out for certain theories. Rohm\myc{Rohm} continues
the discussion of these structures as ``topological defects'' in
the context of quantum gravity.  Rohm also briefly mentions possible
ramifications of exotic structures on the classical Einstein dynamics.
 Bugajska\myc{Bug} pointed out that the
existence of
non-diffeomorphic copies of \R affects the homotopy classification
(kinks) of Lorentzian metrics on the manifold.
\par
On the classical level, can these models really give rise to ``new''
physics, since they are merely new manifolds?  It would certainly seem
so, since the manifold idea underlies all physics.  We might equally ask
whether ``new'' topologies might lead to new physics.  They certainly do,
as evidenced by the impact of ``wormhole'' models, etc. Of course, it
might be argued that there can really be no new physics here, since
standard general relativity can be expressed in terms of arbitrary smooth
manifolds, which necessarily includes these exotic structures.  However,
these exotic manifolds are {\em new}, and have never been explored
before.  They exist as physically distinct manifolds having the same
topology and    this interplay between smoothness and topology has not
been explored in the context of physics before.  \par
  Historically, the progress of theoretical physics has been marked
by increasingly more general relativity principles, involving
increasingly weaker pre-assumptions about reality.   Classical
physics was based on Galilean relativity with its assumption of
absolute time, Maxwell's electromagnetism {\it cum} ether involved
absolute rest.  Einstein generalized these assumptions
with his special and then general theories of
relativity.  The latter  began with the questioning of the necessity
of restricting physics to inertial reference frames (principle of
general relativity), followed by the
questioning of the need for
{\em flatness}, i.e., trivial geometry.  The consequence was
the idea of ``geometry as physics'' with all of the attendant
theoretical structure of general relativity.  Later
in the development of the theory further generalizations
suggested themselves. For example non-trivial topologies, with closed
surfaces not boundaries of volumes, internal symmetries, bundle
and gauge theories, etc.
\par
A common thread in this development is the decreasing set of
{\it unquestioned} mathematical assumptions in the model.  ``Flat'' is
the easiest geometry, but does that mean that nature must
use it?  Must nature use topologically Euclidean space?
Must fields be cross sections of trivial, product bundles?
Corresponding to this decreasing set of assumptions is an increasing
set of mathematical structures available to serve as physical
fields, geometric, topological, and gauge.
\par
 But now,
the surprising discovery  of
\Rx's implies  that there exists an infinity of non-diffeomorphic,
thus {\em physically distinct manifolds}, each with the simple topology
of
\R, but not one of which has yet been investigated as a physical
model!   An interesting exercise   is to imagine what
would have happened if Einstein had used one of these \Rx's in his
early investigation of general, or even special, relativity.  This
conjecture is not realistic, of course, since even today, such manifolds
have not been explicitly constructed in the sense of a coordinate
patch presentation.  Nevertheless,
many of their properties are known, enough hopefully to begin an
investigation of their possible impact on physical theories.
\par
However, we do have  an explicit construction of
exotic structures provided by the Milnor spheres.
  These
might well provide manageable models to investigate explicitly the
influence of exotic smoothness structures on physical theories.  For
example, let \sx\ denote any of the exotic differentiable versions of
the seven-sphere, constructed by Milnor\myc{miln}.
  Standard $S\sp 7$ is of course
the Hopf bundle of Yang-Mills fields over compactified
\R.  The exotic versions are no longer principal $SU(2)$ bundles, but
rather associated bundles with group $SO(4)$.  This fact may
well have significant physical implications.
Also, \sx, may provide some understanding of the differential
geometric restrictions inherent in exotic structures.  For example, it
is clear that no constant curvature complete metric can be put on the
Milnor spheres. The obstruction to continuation of the differential
equations expressing constant curvature can be explicitly analyzed in
these cases, hopefully providing some insight into what may happen in
attempts to continue Einstein metrics in \Rx.\par
Unfortunately, the differential topology involved in studying these
questions is  far from trivial, and mathematical complexities can often
hide  physical simplicities.  Thus, we begin by studying an easily
understood toy model involving complex structures and representations of
the plane vacuum electrostatic equations.      Certainly, we make no
claim that in itself there is any new physics in this model. We only
present it to serve as a readily accessible analog to what might occur in
the case of differentiable structures.\par
\section{Complex Structures as a Toy Model}\myll{cstoy}
     Consider the two-dimensional physics defined by vacuum,  plane
electrostatic  fields,  fully  defined  by  vector  fields  ${\bf
E}(x,y)$
described by component functions $E_x(x,y)$ and $E_y(x,y)$.  The Maxwell
vacuum electrostatic equations are
\beq
{{\partial E_x} \over {\partial x}} +{{\partial E_y} \over {\partial y}}
=0,\myl{cs1}
\beq
{{\partial E_y} \over {\partial x}} -{{\partial E_x} \over {\partial y}}
=0.\myl{cs2}
These are just the Cauchy conditions for the real
and (negative) imaginary parts of  an  analytic  function  of  the
complex variable $z\equiv  x  +  i\  y$.
The most general solution
to \myref{cs1} and \myref{cs2} can be obtained from the complex equation,
\beq
E_x - i\ E_y=f(z),\myl{cs3}
where $f(z)$ is an arbitrary analytic function.
\par
These facts are  well-known and discussed in most
introductory texts on electromagnetism.
Apart from  a  few  illustrative  boundary
value problems, however, they do not seem to lead to any significant
physical
consequences   or   further   insights,   probably   because   the
introduction of a complex structure on the space model is possible
only for two-dimensional problems.   Certainly, it is clear that
the physics is contained in \myref{cs1} and \myref{cs2}, rather than in
\myref{cs3}.
\par
     However,  the  problem  of  {\em complex}  structures  on ${\bf
R}^2$ is
relatively easy, compared to that of the {\em differentiable}  structures
on
${\bf R}^4$ , so we can explicitly  explore  the  relationship  between
the
mathematical structure and its physical implications.   Hence,
for illustrative purposes only, let us assume that the true ``physical''
electrostatic vacuum equations are reduced to \myref{cs3}. We will then
explore what, if any, physical consequences the choice of ``structure'',
complex in this case, has.
\par
     Recall  that  a  complex  structure, $CS$,  on   a   two-dimensional
manifold, $M$, is defined by covering $M$ with  an atlas of charts,
$U_i$ , together  with  maps,  $f_i$   taking $U_i$ (smoothly  and
invertibly) onto open balls in ${\bf R}^2$\ identified with the complex
plane ${\bf C}$ in the ``standard'' way, i.e.,
\beq
(x,y)\varepsilon{\bf R} \leftrightarrow z \equiv x+i\ y,\
z\varepsilon {\bf C}.\myl{cs4}
Furthermore, where defined, $f_i \circ f^{-1}_j$ must be analytic
in ${\bf C}$ in the usual complex sense.  The charts, $U_i$, are
sometimes
called ``coordinate patches'' and, for $p\varepsilon U_i \subset M$,
the value  $z_i\equiv f_i(p)\varepsilon {\bf C}$ is the ``coordinate of p
relative to the patch $U_i$.''  A complex valued function,
$F:V\rightarrow{\bf C}$, for some neighborhood $V\subset M$, is
``analytic'', or ``holomorphic'', if it is complex analytic when
expressed
in the local coordinates, $z_i$, over the $U_i$ covering $V$,  that is,
$F\circ f^{-1}_i$ is analytic (where defined) in the usual sense on
${\bf C}$.
\par
Two such structures on a given $M$, say
$CS^\prime$ given by $\{U^\prime_i,f^\prime_i\}$ and $CS$ given
by $\{U_i,f_i\}$ are {\em equivalent (biholomorphic)} if and only if
there exists a homeomorphism, $F$, of $M$ onto itself such that
$f_i\circ F\circ  f^{\prime-1}_j$ and $f^\prime_i\circ F^{-1}\circ  f^{-
1}_j$
are both holomorphic where defined.  Another way to express
this is to say that the  $CS$ expression of $F$ is analytic
in terms $CS^\prime$ and vice versa.  Note that it is not necessary
that the $CS$ coordinates themselves be analytic in terms of $CS^\prime$,
but only when combined with a homeomorphism.  Thus, let
$U_1=U^\prime_1={\bf R^2}$, with $f_1(x,y)=x+i\ y$, and
$f^\prime_1=x-i\ y$.  Then clearly the primed coordinate is not
analytic in terms of the unprimed one.  However, these are
equivalent complex structures since the homeomorphism,
$F(x,y)=(x,-y)$ satisfies the above condition for equivalence.  That is,
\beq
f_1\circ F\circ f^{\prime-1}_1:z=x+iy\rightarrow (x,-y)\rightarrow
(x,y)\rightarrow x+iy=z.\myl{cs6}
Thus, the physical content of $CS$ and $CS^\prime$ is identical. \par
If $M$ is ${\bf R^2}$ , the  ``standard''  complex
structure, $CS_0$, is defined by the minimal atlas
consisting of only the single $U_1=M$,  and
$f_1(x,y) = x + i\  y$.  We can recast  the  physical  theory  of
two-dimensional vacuum electrostatics by saying that the $x$  and  $y$
components of the electric field must be  the  real  and  (negative)
imaginary parts of an analytic function on $M$, as  defined  by  the
standard complex structure, $CS_0$.  One consequence of this is
that no non-constant vacuum electrostatic field can be bounded.
\par
     However, the standard complex structure  is  not  unique!   There
is
precisely one other inequivalent one.   One  presentation  of  this
second  structure, $CS_1$,  can   be   defined   by   using   some
smooth
diffeomorphism  from  $[0,\infty)$  onto  $[1,0)$,  say  $p$, with
the property that $xp(x^2)$ is bounded.
A simple example is  provided by\ $p(x)=e^{-x}$.
$CS_1$ is then defined by
\beq
(x,y)\rightarrow z_1=p(x^2+y^2)(x+i\ y)\varepsilon {\bf C}.\myl{cs5}
It is now easy to show that $CS_0$ is not equivalent to $CS_1$.  In fact,
if it were then there would exist a function,
$F(x,y)=(F_x(x,y),F_y(x,y))$,
of the plane onto itself such that
$p(F_x(x,y)^2+F_y(x,y)^2)(F_x(x,y)+i\ F_y(x,y))$ would be a global
analytic function
of $x+i\ y$ in the usual sense.  Clearly however this cannot be since
this function is non-constant, but bounded on the entire plane, violating
a well known property of global analytic functions.
\par
Now we can state the physical implications of the choice of structure,
complex in this case: If the physical theory is expressed by the
statement that the $x$ and $y$ components of the electrostatic vacuum
two-dimensional field are real and (negative) imaginary parts of a
function  analytic relative to the chosen complex structure,
then $CS_0$ and $CS_1$ lead to different fields, with physically
measurable differences. In other words,
\state{Physical content of complex structure}{Experiment could
distinguish $CS_0$ from $CS_1.$}  However, experiment cannot distinguish
$CS_0$ from $CS^\prime$ described earlier, since these are biholomorphic.
\par
We repeat that this discussion was intended to be illustrative rather
than of likely physical significance itself.  The description of
electrostatic field theory in terms of analyticity requirements is
certainly not the basis of a general physical theory.  In fact, it could
be argued
that changing the complex structure results in a changed metric and
that the correct theory should include this metric.
However, we
believe that this model provides some motivation for investigating the
possible
physical significance of the choice of {\em differentiable} structures,
whose role in all field theories is indisputable.
\par
 \section{Differentiable structures and Manifolds}\myll{dsm}\par
For convenience, we here review some of the basic definitions and facts
about differential topology\myc{BJ} using the following notation and
terminology.\par
\begin{itemize}
\item
Standard topological ${\bf R^n}$  is defined as the set of     points,
${\bf p}$, each of which can be identified with an n-tuple of real
numbers, $\{p^\alpha\}$. The topology is induced by the usual product
topology of the real line.  For $n=4$, the range of $\alpha$ is
$0,1,2,3.$
\item {\em Topological Manifold}:  This is a topological space which is
locally homeomorphic to ${\bf R^n}.$  We assume all manifolds are
Hausdorff.  For the most part we will be concerned with the $n=4$ case.
\item
{\em Differentiability}:  In this paper, assumed to be $C^\infty$, i.e.,
continuous together with all derivatives.  {\em Smooth} is a synonym for
differentiable in this sense.
\item
A {\em smooth atlas of charts}  on a topological space, $M$, is a
covering of $M$ with open sets, $U_a$, (coordinate patches), together
with maps, ${\bf x_a}$ taking $U_a$ homeomorphically onto an open ball in
${\bf R^n}$.  For each $a$, ${\bf x}_a$ is in fact a n-tuple of real
numbers,
$\{x_a\sp\alpha\}$. Such a pair, $(U_a,{\bf x}_a)$ is called a {\em
chart} and the $\{x_a^\alpha({\bf p})\}$ are the local coordinates of
${\bf p}$ relative to the chart $U_a$. Moreover, where defined, ${\bf
x}_a\circ{\bf x}_b^{-1}$ must
be smooth in terms of the usual ${\bf R^n}$  sense.
\item
A {\em differentiable structure}, $\dM (M)$, is
 a maximal smooth atlas, i.e., the set of all charts compatible with
those of a given smooth atlas.  In the following, we will often define a
$\dM$ by giving one atlas, without referring to the maximalization
process which should be carried out  to induce $\dM$ from a single
member.
\item
A map between two manifolds is {\em smooth} if it is smooth in the usual
real variable sense when expressed in terms of local charts.  A smoothly
invertible smooth map onto is a {\em diffeomorphism.}  Two manifolds are
{\em diffeomorphic} if there is a diffeomorphism between them.  Clearly a
diffeomorphism is a homeomorphism, so diffeomorphic manifolds are
topologically identical.
It is easy to see, as in the example below, that different $\dM$'s can be
placed on a given manifold.  The fundamental question of differential
topology which concerns us in this paper is whether these {\em different}
$\dM$'s are actually {\em diffeomorphic.}
\item
$\Theta$ used as a subscript indicates an {\em exotic},  {\em fake},
or {\em non-standard} differentiable construction.
\item
\Rx is some exotic \R.  Points of \Rx\ will again be labelled by
${\bf p}$, but now $p_\alpha$ are globally defined  continuous functions,
but not globally smooth.  \Rx\ has a $\dMx $\ with coordinate
patches $(U_a,{\bf x_a})$.  The functions $p_\alpha$ restricted to
$U_a$ are continuous functions of $x_a^\alpha$, but cannot be smooth for
all $a$, and this is true for any diffeomorphic copy of $\dMx.$
\item
Finally, we point out that the mathematical notion of diffeomorphism is
normally identified with the physical notion of {\em equivalence under
re-coordinatization}.  Thus, if two manifolds are diffeomorphic, they
provide fully equivalent physical models, simply presented in different
coordinates.
\end{itemize}\par
The concept of differentiable structure is key to the matter of this
paper.  However, it is a subject which can easily be misunderstood.
  In particular, a very natural confusion can arise over the distinction
between the situation in which a given topological  manifold has
different, but diffeomorphic,  differentiable structures and that in
which the different structures are not diffeomorphic.  This is of course
fundamentally important to physical applications since the notion
diffeomorphism is generally taken to signify physical equivalence
associated with ``re-coordinatization.''\par
A counter example can help clarify the somewhat elusive concept of
equivalence and inequivalence of differentiable structures.
  Consider the real line, ${\bf R^1}$,
replacing  the matrix, ${\bf p}$, with its single element, $p$, and
${\bf x}$ with $x$.  The standard structure, $\dM_0$, is generated from
the single global chart with $x=p$.  Relative to $\dM_0$,
 $f(p)$ is thus differentiable if and only if $f(x)$ is in
the usual real variable sense.  Suppose now we  define a ``new''
differentiability structure, $\dM_1,$ by using another global chart with
$u$  as the global coordinate, where $u=p^3$.  Clearly this is acceptable
since $p\rightarrow u$ is a homeomorphism of the manifold onto ${\bf
R^1}$, and there is only one chart.  It is easy to see that these are
indeed {\em different} structures.  If not, their union would also be an
atlas.  But this would require that the transition, $x\circ u^{-1}$ be
smooth.  However, this map takes $y\rightarrow y^{1/3},$ which is not
smooth at the origin.  Thus, $\dM_0\ne\dM_1.$  \par
  However, these two structures are actually equivalent, since
the {\em homeomorphism} $f: p\rightarrow p^{1\over 3}$ of ${\bf R^1}$
onto itself is
a {\em diffeomorphism} of the first structure onto the second. To see
this note that $f$  expressed in local charts becomes $u\circ f\circ x^{-
1}:y\rightarrow (y^{1/3})^3=y,$ the trivial identity map on the
coordinate space ${\bf R^1}$.
In fact, it can be shown rather easily that  any
differentiable structure on ${\bf R^1}$ is equivalent to the standard
one.  Thus, there can be no new physical content to any other $\dM$
on ${\bf R^1},$ nor for {\em any} ${\bf R^n}$ except for the exceptional
case of $n=4.$
\par
\section{Exotic Differentiable structures}\myll{exds}
The first breakthrough in the exploration of  exotic differentiable
structures came in 1956 when Milnor\myc{miln} was able to use an
extension of the Hopf fibering of spheres\myc{st} to construct an exotic
seven-sphere, \sx.  Consider the $S^3$ bundles over $S^4$,
\beq
\begin{array}{lrl}
S^3 & \rightarrow & M^7 \\
&  p & \downarrow \\
& & S^4
\end{array}\myl{b1}
with the rotation group, $SO(4)$, acting on $S^3$, as bundle group.  A
classification of such bundles is provided by $\pi_3(SO(4))\approx Z+Z,$
as discussed in \ \S 18 of    \myc{st}.
This construction can be described in terms of the normal form for $M^7$
in which the base $S^4$ is covered by two coordinate patches, say upper
and lower hemispheres.  The overlap is then ${\bf R^1}\times S^3$ which
has $S^3$ as a retract.  Thus, the bundle transition functions are
defined by their value on this subset, defining a map from $S^3$ into
$SO(4)$ and thus generating an element of $\pi_3(SO(4)).$  The group
action of $SO(4)$ on the fiber $S^3$ can  be conveniently described in
the well-known quaternion form,
\beq
u\rightarrow u'=vu\overline w,\myl{b2}
where $u,v,w$ are all unit quaternions and $\overline w$ is quarternion
conjugate of $w$.  Thus $u\in S^3$ and $(v,w)\in SU(2)\times SU(2)\approx
Spin(4).$ Standard $S^7$ is obtained from the element of $\pi_3(SO(4))$
generated by $(v,1)$, so that the group action reduces to one
$SU(2)\approx S^3$ and the bundle is in fact an $SU(2)$ principal one.
In fact, this is precisely the principle $SU(2)$ Yang-Mills bundle over
compactified space-time, $S^4.$  For more details on classifying sphere
bundles, see
\myc{st},\ \S 20 and, from the physics viewpoint,  \myc{traut}.\par
Milnor's breakthrough in 1956 involved his proving that $M^7$ for the
transition function map, an element of $\pi_3(SO(4)),$ given by
\beq
  u\rightarrow (u^h,\overline u^j)\in Spin(4),\myl{b4}
with $h+j=1$ and $h-j=k,$ and $k^2\ \rlap{$\equiv$}/\ 1 {\rm\ mod\ } 7,$
is in fact exotic, i.e., homeomorphic to $S^7$, but not diffeomorphic to
it.  Clearly, the constructive part is  fairly easy, but the proof of the
exotic nature of the resulting sphere is more involved, drawing from
several important results in differential topology including the Thom
bordism result, cohomology theory, Pontrjagin classes, etc.  Later,
Kervaire and Milnor\myc{KM} and others\myc{gromoll} expanded on these
results, leading to a good understanding of the class of exotic spheres
in dimensions seven and greater. Kervaire and Milnor classified the set
of h-cobordism classes of smooth homotopy n-spheres, which can also be
described as the set of diffeomorphism classes of differentiable
structures on $S^n$.   Moreover, this latter set can also be identified
with the n$^{th}$\ homotopy group of PL/O by smoothing theory for
manifolds. \par
For lower dimensions,  early work by Cerf established that if a smooth
structure on $S^4$ is obtained by gluing two copies of the standard disk
along their $S^3$ boundary by some orientation-preserving diffeomorphism,
then this smooth structure is diffeomorphic to the standard one.  Also,
there is no known example of a smooth compact four-dimensional manifold
whose underlying topological manifold admits only a finite number of
distinct differentiable structures.  On the other hand, bundle theory
ensures that any compact topological n-manifold in all dimensions $n\ge
5$ can have at most a finite number of distinct differentiable
structures.\par
Unfortunately, the path to \Rx is much less easy to  describe\myc{fuff}.
First, we recall that the intersection form of a compact oriented
manifold without boundary, obtained by the Poincare duality pairing of
homology classes in $H_{n-k}$ and $H_k$ can be simply represented in
dimension $n=4=2+2$ by a symmetric square matrix of determinant $\pm 1$.
This form basically reflects the way in which pairs of oriented two-
dimensional closed surfaces fill out the full (oriented) four-space at
their intersection points.  Physicists are perhaps more familiar with
deRham cohomology involving exterior forms for which this intersection
pairing is the volume integral of the exterior product of a pair of
closed two-forms representing the individual cohomology classes, which
again makes sense only in dimension four.  Unfortunately, deRham
cohomology necessarily involves real coefficients and is thus too coarse
for our applications, which need integral homology theory.  At any rate,
this integral intersection
 form, $\omega$,
plays a central role in classifying compact four manifolds.  Whitehead
used it to prove that one-connected closed 4-manifolds are determined up
to homotopy type by the isomorphism class of $\omega.$  Later,
Freedman\myc{fr} proved that $\omega$ together with the Kirby-Siebenmann
invariant classifies simply-connected closed 4-manifolds up to
homeomorphism.  For our purposes, the important result was that there
exists a topological four manifold\footnote{We use the standard notation
in
which $\vert\omega\vert$ is a topological manifold having $\omega$ as
intersection form.}, $\vert E_8\vert,$ having intersection form
$\omega=E_8,$ the Cartan matrix for the exceptional lie algebra of the
same name.  As it stands, Freedman's work is in the topological category,
and does not address smoothness questions.  The theorem of Rohlin\myc{R}
states that the signature of a closed connected oriented smooth 4-
manifold must be divisible by 16, so that $\vert E_8\vert$ cannot exist
as a {\em smooth} manifold since its signature is 8.   Next, Donaldson's
theorem\myc{D} provides the crucial (for our purposes) generalization of
this result to establish that
$\vert E_8\oplus E_8\vert$ is not smoothable, even though its signature
is 16. The work of Donaldson is based on the moduli space of solutions to
the $SU(2)$ Yang-Mills equations on a four-manifold, which first occur in
physics literature.\par
Having established some algebraic machinery, the next step involves an
algebraic variety, the Kummer surface, $K$, a real four-dimensional
smooth manifold in $CP^3.$  It is known that
\beq
K=\vert -E_8\oplus -E_8 \oplus 3\pmatrix{ 0 & 1 \cr 1 & 0}\vert.\myl{k1}
The last part of this intersection form is easily seen to be realizable
by $3(S^2\times S^2)$, which is smooth.  Thus, Donaldson's theorem
implies that it is impossible to do smooth surgery on $K$ in just such a
way as to excise the smooth $3(S^2\times S^2)$, leaving a smooth
(reversing orientation) $\vert E_8\oplus E_8\vert$. In the following, we
refer to these two parts as $V_1$(smoothable) and $V_2$(not smoothable)
respectively, so smooth $K=V_1\cup V_2.$ In investigating the failure of
this smooth surgery Freedman found the first fake \Rx.  Using a
topological $S^3$ to separate $V_1$ from $V_2$, Donaldson's result showed
that this $S^3$ cannot be smoothly embedded, since otherwise $V_2$ would
have a smooth structure.  However, by further surgery, it is found that
this dividing $S^3$  is also topologically embedded in a topological \R\
and actually includes a compact set in its interior. Thus we have\par
\state{Existence of exotic \Rx} {This topological \R contains a compact
set which cannot be contained in any smoothly embedded $S^3.$  This
surprising result then implies that this manifold is indeed an \Rx\ since
in any diffeomorphic image of
\R
 every compact set is included in the interior of a smooth sphere.}\par
Since then, there have been many developments, some of which are
summarized in the book by Kirby\myc{surv}.  Unfortunately, none of the
uncountable infinity of \Rx's has been presented in explicit atlas of
charts form, so most of the properties can only be described indirectly,
through existence or non-existence type of theorems.\par
For example, some information about differential geometry on such a
manifold can been
obtained, such as\par
{\bf Theorem 1:}  There can be no  geodesically complete
metric (of any signature) with non-positive sectional curvature on
\Rx.\par
Proof:  If there were such a metric, the Hadamard-Cartan theorem could
be used to show that the exponential map would provide   a diffeomorphism
of
the tangent space at a point onto \Rx.
   In particular, there can be no flat geodesically complete metric. For
more discussion on exotic geometry, see \myc{Reinhart}.
Natural questions then arise concerning the nature of the obstructions
to continuing the solutions to the differential equations expressing
flatness in the natural exponential coordinates.  In physics,
obstructions to continuation of solutions are often of considerable
significance, e.g., wormhole sources.  However, up to now, such
obstructions generally have been  a result of
either  topology, (incompleteness caused by excision),
or some sort of curvature singularity.  Neither of these is present here.
This
problem is particularly interesting for those \Rx 's which cannot be
smoothly embedded in standard \R, which thus cannot be geodesically
completed with a flat metric.  \par
Another useful result is \par
{\bf Theorem 2:}  There exists a smooth copy of each \Rx\ for which the
global $C\sp 0$ coordinates are smooth in some neighborhood.  That is,
there exists a smooth copy,
${\bf R\sp 4}_\Theta=\{(p^\alpha)\}$,
for which $p^\alpha
\in      C\sp \infty$ for $\vert {\bf p}\vert<\epsilon.$
\par
Proof:  This may be obvious, but we seem to need a rather involved
argument using the Annulus Theorem\myc{annulus}.\par
What this gives is a local smooth coordinate patch, on which standard
differential geometry can be done, but which cannot be extended
indefinitely.  The obstruction should be physically interesting.
Also, this theorem leads naturally to the following construction.\par
By puncturing \Rx, we get a ``semi-exotic'' cylinder, i.e,
$\mRx-\{0\}\simeq {\bf R\sp 1}\xt S\sp 3,$  where $\xt$ means topological
but not smooth product.  By ``semi-exotic'' we mean that the product is
actually smooth for a semi-infinite extent of the first coordinate.
This might be a very interesting cosmological model for physics, which
after the big bang is ${\bf R}\sp 1 \times S\sp 3. $   Here we would
run into an obstruction to continuing the smooth product structure at
some finite time (first coordinate) for some unknown,
 but potentially very interesting, reason.\par   An
even more interesting possibility to consider would involve
localizing the ``fakeness'' in some sense.  One version that
comes to mind would happen if we could smoothly glue two such
semi-exotic cylinders at their exotic ends.  Of course a second gluing
at their smooth ends would then give an exotic smoothness on the
topological product,  $S\sp 1 \times S\sp 3$. The existence of such
an   $S\sp 1\xt S\sp 3$ is not known, as far as we know. We proceed to
summarize a set of conjectures.\par
\section{Conjectures}\myll{conj}\par
What are the
possible physical implications of the existence of the exotic
spaces?  First, consider the \sx,
which can be explicitly constructed.  Perhaps they could be considered
as possible models for  exotic Yang-Mills theory. Some \sx\ are
 $SU(2)$ bundles, but not  principle ones, since their groups must
be $SO(4)$.  This would contrast with standard Yang-Mills
structure\myc{yama} in which the total space is $S^7$ regarded as a
principle $SU(2)$ bundle.  Next, \sx\ can be used as  toy space-time
models,
serving as the base manifolds for various geometric and other field
theories.  Various questions of physical interest can then be asked
on these models and the answers compared to those obtained from standard
$S^7$.  For example, the non-existence of a constant curvature metric
on \sx\ has already been thoroughly explored\myc{gromoll}.  The analysis
of such differential geometric problems on \sx\ as compared to $S^7$
 should give some indication of the type of results
that could come from physics on \Rx\ as compared to that on standard
${\bf R^4}.$
\par
There are also questions concerning the
physical implications of doing general relativity on \Rx.
First, several questions of physical significance but  of
 a more mathematical nature come to mind:\par
{\bf Question:}  Does there exist an \Rx\  which is standard outside a
compact set?  \par Thus, in the above notation,  there would be a copy
for which the global continuous coordinates are smooth outside a
sphere, i.e.,
$p^\alpha \in C^\infty$ for $\vert{\bf p}\vert>k$.  Clearly this
would be an inversion of the result above, Theorem 2.
  Of course, this is likely to
be a very difficult question since if such an \Rx\ were found,
exotic structures for many 4-manifolds could be obtained from given
smooth structures by using the \Rx\ as a chart in a new atlas.
  Perhaps an easier question, but one of even greater
physical significance would be the following:\par  {\bf Question:} Does
there exist an \Rx\ for which the global continuous coordinates are
smooth outside of a cylinder, i.e., $p^\alpha \in C^\infty$ for $p^0>0$
and $(p^1)^2+(p^2)^2+(p^3)^2> k?$\par
Physically, such a structure could provide an interesting model for the
world line of a particle.  At spatial infinity, everything is standard,
geometry
can be flat, but this flat geometry cannot be continued into the world
line at the origin.
This is basically the way particle sources occur in general relativity.
A variation on this, still of physical interest is \par {\bf Question:}
Does there exist an \Rx\ for which the $p^\alpha \in C^\infty$ for
$p^0<0?$ \par This could be of physical interest if $p^0$ is time,
so that the model is standard for semi-infinite time, but cannot be
continued this way indefinitely.\par
{\bf Question:} What is the nature of Cauchy development in light of the
existence of \Rx?
 Specifically, does there exist a closed
smooth ${\bf R^3}$ in
\Rx ?\par
  Actually, the entire problem of developing a manifold from a coordinate
patch piece on which an Einstein metric is known, still has many
unanswered
aspects.  Recall for example the evolution of our understanding of the
appropriate manifold to support the (vacuum)
 Schwarzschild metric. Originally,
the solution was expressed using $(t,r,\theta,\phi)$ coordinates
as differentiable outside of the usual ``coordinate singularities''
well known for spherical coordinates.  However, the Schwarzschild
metric form itself in these coordinates exhibits another singularity
on $r=2m$, sometimes referred to as the ``Schwarzschild singularity.''
Later work, culminating in the Kruskal representation, showed that
the Schwarzschild singularity could be regarded as merely another
coordinate one  in the same sense as is the z-axis for $(r,\theta,\phi)$.
This example helps to illustrate that
 in general relativity our understanding of the
physical significance of a particular metric often undergoes an evolution
as various coordinate representations are chosen.  In this process,
the topology and differentiable structure of the underlying manifold
may well change.  In other words, as a practical matter, the study of the
completion of a locally given metric often involves the construction of
the global manifold structure in the process.  Could any conceivable
local Einstein metric lead to  an \Rx\ by such a process?
\par
Of course, local coordinate patch behavior is of great importance
to physics, so another set of physically interesting questions would
relate to the coordinate patch study of \Rx.   This may be
too difficult of a task for present mathematical technology, but  some
questions may be reasonable.
For example, can some \Rx's be covered by only a finite
number of coordinate patches?  If so, what is the minimum number?
  What are the intersection
properties of the coordinate patch set which makes it non-standard?\par
A directly physical set of questions to be considered would stem
from attempts to embed known solutions to the Einstein equations in
\Rx, then asking what sort of obstruction intervenes
to prevent their indefinite, complete, continuation, in this space.
A particular example would be the embedding of a standard
homogeneous and isotropic cosmological
metric in ${\bf R^1}\xt S^3$ discussed above.  Clearly the isotropy
cannot be continued indefinitely.  Why?  What is the physical
significance
of this obstruction?  \par
\section{Conclusion}\myll{conc}
{}From the time of Einstein, the importance of separating physically
{\em invariant} statements from those that depend on the choice
of the observer has been generally recognized in principle, if
not in practice.  Mathematically, this means first taking care
to identify those transformation groups in the mathematical
model of a theory with the physical operation of performing
an (idealized) coordinate transformation. Only statements
whose validity is invariant under these transformation groups
can then be regarded as having absolute physical significance
and testability.  Of course, in practice, it is often convenient
to restrict a particular argument to a subclass of these
transformations, but, in principle, this restriction should
be kept in mind.  In most applications, these transformations
must be {\em smooth}, so investigation into smoothness properties of
given topological manifolds surely has physical significance.
\par
Finally, it is indeed true that the existence of \Rx's does not in any
way change the {\em local} physics of general relativity or any other
field theory.  However, it has long been known that global questions can
have profound effects on a physical theory.  Until recently, physicists
have thought of global matters almost exclusively as being of purely {\em
topological} significance, whereas we now know that at least in the
physically important case of ${\bf R^4},$ there are very exciting global
questions related to differentiability structures, the way in which local
physics is patched together smoothly to make it global.  Certainly, the
\Rx's are essentially just ``other'' manifolds.  However, there are an
infinity of them which have never been remotely considered in the
physical context of classical space-time physics on Einstein's original
model, \R.  It would be surprising indeed if none of these had any
conceivable physical significance.

\end{document}